\documentclass[10pt,letterpaper]{article}
\usepackage[latin9]{inputenc}
\usepackage{color}
\usepackage{units}
\usepackage{amsmath}
\usepackage{amssymb}
\usepackage{graphicx}
\usepackage[unicode=true,
 bookmarks=false,
 breaklinks=false,pdfborder={0 0 1},backref=false,colorlinks=true]
 {hyperref}
\hypersetup{
 citecolor=blue,urlcolor=blue}

\makeatletter

\pdfpageheight\paperheight
\pdfpagewidth\paperwidth

\usepackage{osameet3}%% use version 3 for proper copyright statement

\usepackage{cite}
\usepackage[printonlyused]{acronym}
\acrodef{DM}{distribution matcher}
\acrodef{Hi-DM}{hierarchical DM}
\acrodef{PS}{probabilistic shaping}
\acrodef{PAS}{probabilistic amplitude shaping}
\acrodef{AWGN}{additive white Gaussian noise}
\acrodef{QAM}{quadrature amplitude modulation}
\acrodef{FEC}{forward error correction}
\acrodef{PAM}{pulse amplitude modulation}
\acrodef{LUT}{look up table}
\acrodef{MB}{Maxwell-Boltzmann}
\acrodef{ESS}{enumerative sphere shaping}
\acrodef{CCDM}{constant composition distribution matching}
\acrodef{IR}{information rate}
\acrodef{SNR}{signal to noise ratio}
\acrodef{PPM}{pulse position modulation}
\acrodef{invDM}{inverse DM}
\acrodef{TX}{transmitter}
\acrodef{RX}{receiver}
\acrodef{BER}{bit error rate}
\acrodef{SER}{symbol error rate}

\makeatother

\begin{document}

\title{Hierarchical Distribution Matching: a Versatile Tool for Probabilistic
Shaping}

\author{Stella Civelli$^{1,2}${*}, Marco Secondini$^{1,2}$}

\maketitle
\address{$^1$ TeCIP Institute, Scuola Superiore Sant'Anna, Via G. Moruzzi 1, 56124, Pisa, Italy\\$^2$ PNTLab, CNIT, Via G. Moruzzi 1, 56124, Pisa, Italy}
\email{{*}stella.civelli@santannapisa.it}

%% Uncomment the following line to override copyright year from the default current year.
\copyrightyear{2020}
\begin{abstract}
The hierarchical distribution matching (Hi-DM) approach for probabilistic
shaping is described. The potential of Hi-DM in terms of trade-off
between performance, complexity, and memory is illustrated through
three case studies. 
\end{abstract}

\section{Introduction}

\vspace{-4pt}

Recently, \ac{PS} techniques have been widely investigated to improve
the performance and the flexibility of optical fiber networks. By
assigning different probabilities to the constellation symbols---e.g.,
trying to approximate the capacity-achieving Gaussian distribution
for the \ac{AWGN} channel---\ac{PS} allows both to finely adapt
the \ac{IR} to the available \ac{SNR} and to reduce the gap between
the \ac{IR} achievable with uniform QAM constellations and the channel
capacity \cite{buchali2016JLT,fehenberger2016JLT}.

An effective \ac{PS} approach, named \ac{PAS} and based on a proper
concatenation of a fixed-to-fixed-length \ac{DM} and a systematic
\ac{FEC} encoder, has been proposed in \cite{bocherer2015bandwidth}.
The key element of PAS is the \ac{DM}, which maps $k$ uniformly
distributed bits on $N$ shaped (according to a desired distribution)
amplitudes from the alphabet $\mathcal{A}=\left\{ 1,3,\dots,2M-1\right\} $
with rate $R=k/N.$ This map induces a specific structure on the output
sequence, whose elements are, therefore, not independent. Consequently,
the rate $R$ of the \ac{DM} is lower than the entropy rate $\mathcal{H}$
that would be obtained with a sequence of i.i.d. amplitudes with the
same target distribution, yielding the rate loss $R_{loss}=\mathcal{H}-R\geq0.$

Different methods to realize a \ac{DM} with a low rate loss have
been recently proposed \cite{fehenberger2018multiset,schulte2016CCDM,yoshida2019hierarchicalDM,gultekin2018Sphereshaping}.
While the rate loss usually tends to zero when the block length $N$
increases (but with different convergence speed for different DMs),
this happens at the expense of increased computational cost, memory,
and/or latency. To address this issue, an effective solution based
on a \ac{Hi-DM} structure based on \ac{LUT}s has been recently proposed
\cite{yoshida2019hierarchicalDM,yoshida2019preferred}. Elaborating
on the latter idea, our work investigates a generalized \ac{Hi-DM}
approach in which several short \ac{DM}s are hierarchically combined
to obtain a longer \ac{DM} with reduced rate loss and limited complexity,
memory, and latency. The proposed approach can be tailored to look
for the best trade-off between the mentioned parameters, depending
on the system requirements. To show the potential of Hi-DM, we illustrate
three different structures.

\section{Hierarchical distribution matcher\label{sec:HiDM}}

\vspace{-4pt}

\begin{figure}[htbp]
\centering \includegraphics[width=0.95\columnwidth]{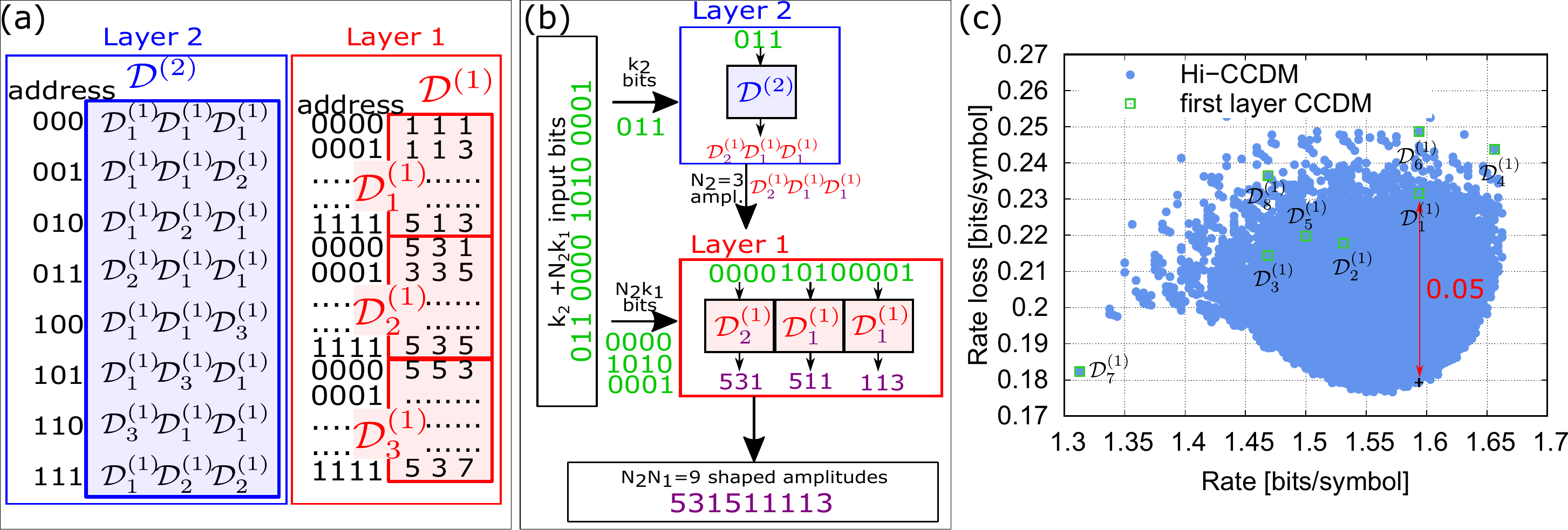}\caption{\label{fig:HiDM}Example of (a) Hi-DM structure and (b) encoding with
2 layers of \ac{LUT}, $\mathbf{N}=(3,3)$, $\mathbf{k}=(4,3)$, and
$\mathbf{M}=(4,3)$. (c) Rate loss versus rate for a $2$-layer Hi-DM
using CCDM on both layers.}
\end{figure}

The \ac{Hi-DM} has a layered structure. The lowest layer consists
of a set of short \emph{disjoint} (i.e., without common output sequences)
\ac{DM}s that encode the input bits on the output levels. In turn,
this set of \ac{DM}s is seen by the upper layer as a \emph{virtual
alphabet}. Thus, in the upper layer, another \ac{DM} is used to encode
some additional input bits on the particular sequence of \ac{DM}s
that will be used in the lower layer. If more layers are desired,
a set of disjoint \ac{DM}s (rather than a single \ac{DM}) is considered
also in the upper layer and an additional layer is added. Compared
to the single short DMs, the overall structure has a longer overall
block length and more degrees of freedom in the generation of the
output sequences with the desired target distribution, yielding a
reduction of the rate loss with a moderate increase of computational
and memory resources. 

The working principle is illustrated by considering the 2-layer \ac{Hi-DM}
example in Fig.~\ref{fig:HiDM}(a). The upper layer (layer 2) is
characterized by a \ac{DM} $\mathcal{D}^{(2)}$ that takes $k_{2}=3$
input bits and maps them to $N_{2}=3$ output symbols. Here, the output
symbols are not ``conventional'' amplitude levels, but are taken
from an alphabet of $M_{2}=3$ \ac{DM}s, $\mathcal{D}^{(1)}=\{\mathcal{D}_{1}^{(1)},\mathcal{D}_{2}^{(1)},\mathcal{D}_{3}^{(1)}\}$.
The output of layer 2 corresponds to the specific sequence of \ac{DM}s
to be used in layer 1. In turn, layer 1 is characterized by the set
of \ac{DM}s $\{\mathcal{D}_{1}^{(1)},\mathcal{D}_{2}^{(1)},\mathcal{D}_{3}^{(1)}\}$,
each with $k_{1}=4$ input bits and $N_{1}=3$ output levels taken
from the $M_{1}$-ary alphabet $\{1,3,\ldots2M_{1}-1\}$. Thus, layer
1 takes $N_{2}k_{1}=12$ input bits and eventually generates $N_{2}N_{1}=9$
output levels. An example of encoding is shown in Fig.~\ref{fig:HiDM}(b).
The use of disjoint \ac{DM}s in layer 1 is required to ensure that
the encoding process can be inverted at the decoder.

With respect to considering a predetermined sequence of DMs in layer
1, which would simply result in a rate $R_{1}=k_{1}/N_{1}$, the proposed
structure allows to encode ``for free'' $k_{2}$ additional bits
on the same number of output levels, increasing the overall rate to
$R=(k_{2}+N_{2}k_{1})/(N_{2}N_{1})>R_{1}$ and potentially reducing
the rate loss. 

The \ac{Hi-DM} structure can be extended to $L$ layers and is characterized
by the vectors $\mathbf{N}=(N_{1},\ldots N_{L})$, $\mathbf{k}=(k_{1},\ldots,k_{L})$,
and $\mathbf{M}=(M_{1},\ldots,M_{L})$. Each layer can use any kind
of DM (e.g., \ac{CCDM} \cite{schulte2016CCDM}, \ac{ESS} \cite{gultekin2018Sphereshaping},
or \ac{LUT}) and, given the properties of the lower layer, can be
designed to target the desired distribution. In the following examples,
the \ac{Hi-DM}s structures are designed to minimize the mean energy
per symbol for a given rate or, equivalently, to target a Maxwell--Boltzmann
distribution \cite{fehenberger2016JLT}.

\section{Three case studies: structure and performance}

\begin{figure}[htbp]
\centering \includegraphics[width=0.8\columnwidth]{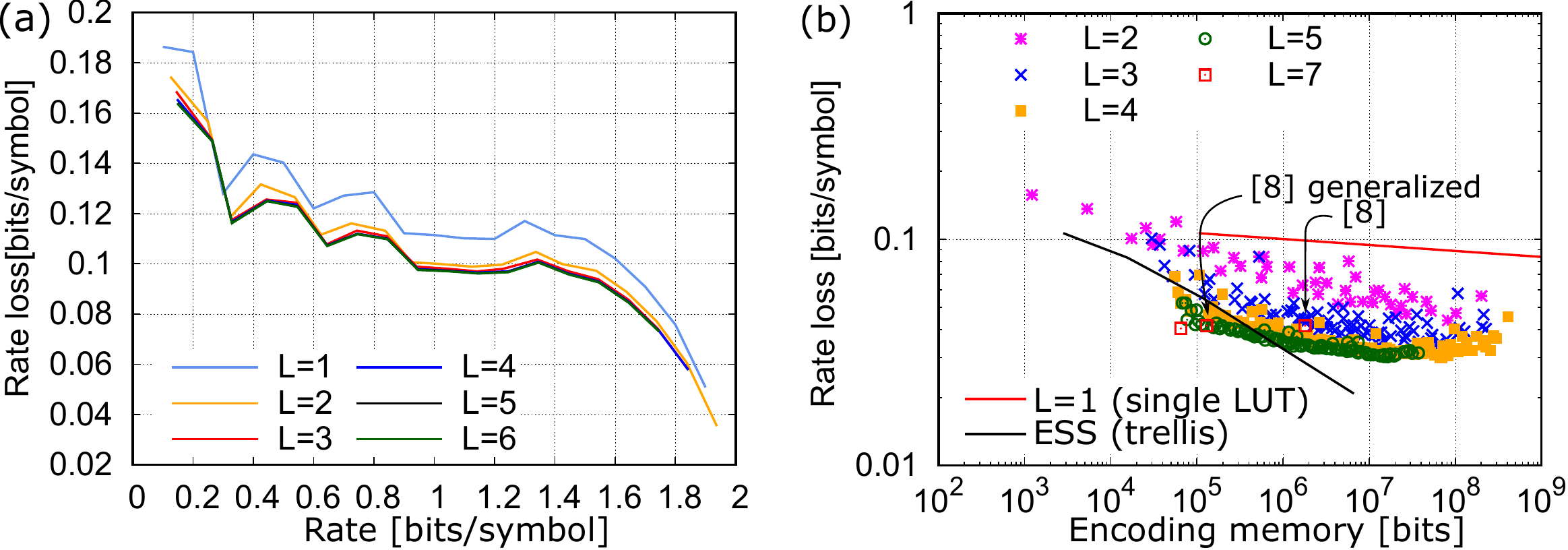}\caption{\label{fig:results}Rate loss (a) versus rate for Hi-DM using PPM
on all the layers except the first, and (b) versus encoding memory
for Hi-DM using LUTs on all the layers.}
\end{figure}

\vspace{-10pt}The Hi-DM structure can be used to combine different
\ac{CCDM} on several layers. In this case, the Hi-DM remains fixed-to-fixed
length even if the DMs on the same layer have a different number of
input bits. The potential of this method is illustrated in Fig.\ \ref{fig:HiDM}(c),
which shows the rate loss versus rate that can be obtained with a
$2$-layer Hi-DM combining $M_{2}=8$ \ac{CCDM}s on layer~1 (block
length $N_{1}=32$, $M_{1}=4$ amplitudes, different rates; $\mathcal{D}_{1}^{(1)}$
optimized for the target rate of $1.59$, the others for different
rates)---whose rate loss is reported with green squares---with an
external \ac{CCDM}, with block length $N_{2}=10$. The blue dots
show the rate loss versus rate that can be obtained with different
\ac{CCDM} compositions on the second layer. At the target rate, the
rate loss decreases of about 23\%, from $0.2316$ for the 1-layer
\ac{CCDM} to $0.1793$ for the $2$-layer Hi-DM.

In a second example, we employ the \ac{PPM} concept to decrease the
rate loss with a negligible complexity. Given a set of $L$ disjoint
DMs with increasing energy on the first layer, the external layers
$\ell=2,\dots,L$ are used according to a \ac{PPM} scheme as follows.
The second layer selects $N_{2}-1$ times the lower energy DM $\mathcal{D}_{1}^{(1)}$,
and once $\mathcal{D}_{2}^{(1)}$, encoding $\log_{2}(N_{2})$ bits
on the \emph{position} of the latter. The third layer applies $N_{3}-1$
times the second layer, and once the same structure but replacing
$\mathcal{D}_{2}^{(1)}$ with $\mathcal{D}_{3}^{(1)}$, encoding $\log_{2}(N_{3})$
bits on its position. This method increases the rate, with only a
slight increase of the energy and a negligible additional complexity.
The Hi-DM with \ac{PPM} is investigated considering on the first
layer $L$ \ac{LUT}s with block length $N_{1}=10$, variable $k_{1}$,
and $M_{1}=4$ amplitudes. The first LUT $\mathcal{D}_{1}^{(1)}$
contains the first $2^{k_{1}}$ minimum energy sequences, the second
LUT $\mathcal{D}_{2}^{(1)}$ contains the second $2^{k_{1}}$ sequences
with minimum energy, and so on. Fig.\ \ref{fig:results}(a) shows
the rate loss (the minimum for each $k_{1}$) for different rates
and a different number of layers ($L=1$ means that only the LUT $\mathcal{D}_{1}^{(1)}$
is used, without Hi-DM). For a given rate, the rate loss decreases
when increasing the number of PPM layers, converging to the minimum
value with approximately $L=3$ layers. As expected, the rate loss
vanishes when the rate tends to $2$ (uniform constellation), for
which no shaping is performed. At most, the rate loss reduction with
respect to the single LUT slightly exceeds $0.015$, with an improvement
of about $17\%$ at rate $1.6$.

The proposed Hi-DM structure can use \ac{LUT}s on all the layers,
designing each LUT to minimize the mean output energy for a given
rate. In this case, it becomes practically equivalent to the structure
considered in \cite{yoshida2019hierarchicalDM,yoshida2019preferred}.
With LUTs, computational complexity is usually negligible and memory
becomes the most important resource. The Hi-DM with LUTs is investigated
at the target rate $R=507/320\approx1.58$ as in \cite{yoshida2019hierarchicalDM,yoshida2019preferred}.
Fig.\ \ref{fig:results}(b) shows the rate loss versus encoding memory
with a different number of layers. The memory is evaluated as the
number of bits required to store all the LUTs. A possible parallelization
of the whole structure or of some of the inner LUTs (as implicitly
done in \cite{yoshida2019hierarchicalDM,yoshida2019preferred}) may
be considered, depending on the system requirements. The figure also
reports the memory required by a single LUT approach (red line), and
the memory required to store the trellis structure for the \ac{ESS}
approach \cite{gultekin2018Sphereshaping} (black line) which, however,
has a relevant computational complexity. Fig.\ \ref{fig:results}(b)
shows some of the values with the best trade-off between rate loss
and memory for $L\leq5$, while for $L=7$ only three values are shown,
two of which correspond to the structure proposed in \cite{yoshida2019hierarchicalDM,yoshida2019preferred},
without the implicit parallelization (i.e., implemented according
to our structure and labeled ``\cite{yoshida2019preferred} generalized'')
and as given by the authors (labeled ``\cite{yoshida2019preferred}'').
The figure shows that increasing the number of layers, the rate loss
significantly decreases with the same overall memory, or, equivalently,
the memory required for a certain rate loss can be substantially diminished.
Finally, the figure shows that the performance (in terms of rate loss
versus memory) of the structure proposed in \cite{yoshida2019hierarchicalDM,yoshida2019preferred}
can be even improved with some structures.

\section{Conclusion}

\vspace{-4pt}

We presented a general Hi-DM structure for the effective implementation
of \ac{PAS}. The proposed Hi-DM structure can be tailored according
to the system requirements, including target distribution, available
hardware resources, and rate loss. We have shown the potential of
Hi-DM through three simple examples: i) a two-layer combination of
short CCDMs, which reduces the rate loss by approximately $23\%$
compared to a single CCDM with same block length; ii) a combination
of a LUT-based sphere shaping layer with several PPM-like layers,
which reduces the rate loss of the LUT-based DM by $10\%$ with a
negligible additional complexity; and iii) an $L$-layer combination
of LUTs, in which the rate loss can be reduced by increasing the number
of layers while keeping the required LUT memory small. In the last
example, the LUT-based Hi-DM achieves a rate loss of about $0.04$
(for the rate $1.58$ on $4$ amplitude levels) with less than $\unit[100]{Kbit}$
of memory.

\section*{Acknowledgment}

\vspace{-4pt}

This work was supported in part by ENYGMA. We thank T. Yoshida
for sharing the details of the structure in \cite{yoshida2019hierarchicalDM,yoshida2019preferred}.

\bibliographystyle{osajnl}
\bibliography{ref}

\end{document}